\documentclass[twocolumn,english]{revtex4}
\usepackage[T1]{fontenc}
\usepackage[latin9]{inputenc}
\usepackage{array}
\usepackage{amstext}
\usepackage{amssymb}
\usepackage{graphicx}


\usepackage{babel}
\begin{document}

\title{Stability analysis of a model gene network links aging, stress resistance, and negligible senescence}

\author{Valeria Kogan$^{1,2}$, Ivan Molodtcov$^{1,2}$, Leonid I. Menshikov$^{2,3}$,
Robert J. Shmookler Reis$^{4,5,6}$ and Peter Fedichev$^{1,2}$}

\address{$^{1)}$Moscow Institute of Physics and Technology, 141700, Institutskii
per. 9, Dolgoprudny, Moscow Region, Russian Federation}

\address{$^{2)}$Quantum Pharmaceuticals, Ul. Kosmonavta Volkova 6A-606, 125171,
Moscow, Russian Federation}

\address{$^{3)}$Northern (Arctic) Federal University, 163002, Severnaya Dvina Emb. 17, Arkhangelsk, Russian Federation}

\address{$^{4)}$McClellan VA Medical Center, Central Arkansas Veterans Healthcare
System, Little Rock, AR, USA}

\address{$^{5)}$Department of Biochemistry and Molecular Biology, University
of Arkansas for Medical Sciences, Little Rock, AR, USA}

\address{$^{6)}$Department of Geriatrics, University of Arkansas for Medical
Sciences, Little Rock, AR, USA}
\begin{abstract}
Several animal species are considered to exhibit what is called negligible
senescence, i.e. they do not show signs of functional decline or any
increase of mortality with age, and do not have measurable reductions
in reproductive capacity with age. Recent studies in Naked Mole Rat (NMR) 
and long-lived sea urchin showed that
the level of gene expression changes with age is lower than in other organisms.
These phenotypic observations correlate well with exceptional
endurance of NMR tissues to various genotoxic stresses. Therefore,
the lifelong transcriptional stability of an organism may be a key
determinant of longevity. However, the exact relation between genetic
network stability, stress-resistance and aging has not been defined.
We analyze the stability of a simple genetic-network model
of a living organism under the influence of external and endogenous
factors. We demonstrate that under most common circumstances a gene
network is inherently unstable and suffers from exponential accumulation
of gene-regulation deviations leading to death. However, should
the repair systems be sufficiently effective, the gene network can
stabilize so that gene damage remains constrained along with mortality
of the organism, which may then enjoy a remarkable degree of stability
over very long times. We clarify the relation between stress-resistance
and aging and suggest that stabilization of the genetic network may
provide a mathematical explanation of the Gompertz equation describing 
the relationship between age and mortality in many species, and of the
apparently negligible senescence observed in exceptionally long-lived 
animals. The model may support a range of applications, such as 
systematic searches for therapeutics to extend lifespan and healthspan.
\end{abstract}


\maketitle

The Naked Mole Rat is an example from a growing list of animal species
with no signs of aging or reduction in reproductive capacity with age. On the 
other hand, the age-dependent increase in death rate for most species, 
including humans, follows the Gompertz equation, which describes an 
exponential increase in mortality with age. In this work we construct 
a mathematical model of a gene network, described by a few differential 
equations with unstable and stable solutions, corresponding to normal 
and ``neglible'' senesecence.  For this purpose, we define aging as an 
exponential accumulation of epigenetic dysregulation errors, and show 
that the Gompertz ``law'' is a direct consequence of the inherent 
genetic instability of the network. On the organism level this leads 
to the loss of stress resistance, the onset of age-related diseases, 
and finally to death. We suggest and analyze several strategies for 
gene network stabilization, which can be exploited for future life-extending 
therapeutics.

\section*{Introduction}

Aging in most species studied, including humans, leads to an exponential
increase of mortality with age, primarily from a variety of age-related
diseases. A growing number of animal species are recognized to exhibit
what is called negligible senescence, i.e. they do not show measurable
reductions with age, in their reproductive ability or functional 
capacities \cite{finch1994longevity}. Death rates in negligibly senescent
animals do not increase with age as they do in senescent organisms.
One negligibly senescent species is the ocean quahog clam, which
lives about $400$ years in the wild \cite{abele2008imperceptible}
and is the longest-living non-colonial animal. Its extreme longevity
is associated with increased resistance to oxidative stress in comparison
with short-lived clams \cite{ungvari2011extreme}. No noticeable signs
of aging were found in a few turtle species, such as Blanding's turtle,
whose lifespan is over 75 years \cite{congdon2001hypotheses}, and the
painted turtle, which was documented to live at least 61 years. Studies
showed that these turtles increase offspring quality with age, so they
are considered to be negligibly senescent \cite{congdon2003testing}.
The archetypical example of negligible senescence is the Naked Mole
Rat (NMR), which has been documented to live in captivity for as
long as 28 years \cite{buffenstein2005naked} with no signs of increasing
mortality, little or no age-related decline in physiological functions, 
sustained reproductive capacity, and no reported instances of cancer 
throughout their long lives \cite{edrey2011endocrine}. These phenotypic
observations correlate well with exceptional resistance of NMR tissues
to diverse genotoxic stresses \cite{lewis2012stress,andziak2006high}.
Even more examples of negligibly senescent organisms may be found
in the AnAge database \cite{tacutu2013human}.

In contrast, aging in most species studied, including humans, follows 
the Gompertz equation \cite{gompertz1825nature} describing an 
exponential increase of mortality with age, a possible signature of 
underlying instability of key regulatory networks. Recent studies of 
gene expression levels in the Naked Mole Rat and long-lived sea urchin 
\cite{ebert2008longevity} showed that the frequency of gene expression 
changes is lower in negligibly senescent animals than in other animal
species \cite{loram2012age-related,kim2011genome,ebert2008longevity,ungvari2011extreme,abele2008imperceptible,congdon2001hypotheses}.
Therefore, the lifelong stability of the transcriptome may be a key
determinant of longevity, and improving the maintenance of genome stability may be a sound
strategy to defend against numerous age-related diseases. In this
work we propose and analyze a simple mathematical model of a genetic
network, and investigate the stability of gene expression levels in
response to environmental or endogenous stresses. We show that under
a very generic set of assumptions there exist two distinctly different
classes of aging dynamics, separated by a sharp transition depending
on the genome size, regulatory-network connectivity, and the efficiency
of repair systems. If the repair rates are sufficiently high or the
connectivity of the gene network is sufficiently low, then the regulatory
network is very stable and mortality is time-independent in a manner
similar to that observed in negligibly senescent animals. Should the
repair systems display inadequate efficiency, a dynamic instability
emerges, with exponential accumulation of genome-regulation errors,
functional declines and a rapid aging process. The two regimes also
show dramatically different dynamics of stress-resistance with age:
stable genetic networks are more robust against noise, and the efficacy
of stress defenses does not decline with age. In contrast, the stress
responses of ``normally aging'' animals deteriorate exponentially.



\section*{Genetic-network stability analysis}
\begin{figure}
\includegraphics[width=0.99\columnwidth]{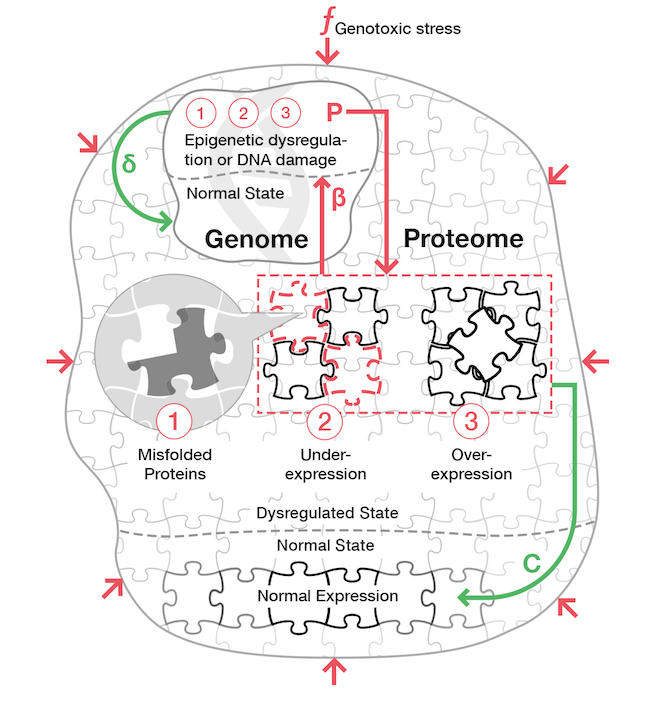}

\caption{The minimum stability analysis model for a gene network. At any given time
the genome consists of a number of normally expressed  and dysregulated
genes . The proteome accumulates ``defects'', the proteins
expressed by dysregulated genes, which are removed via the protein 
quality-control or turnover systems. DNA repair machinery
controls epigenetic states of the genes and restores normal
expression levels. On top of this, interactions with the environment
damage both the proteome and the genome subsystems, increasing the
load on the protein-turnover and DNA-repair components. Parameters
$f$, $\beta$, $p$ and $c$ appear in Eqs. (\ref{eq:basic_b}) and 
(\ref{eq:basic_M}), and are interpreted below.}

\end{figure}

A living body, the organism, is an interacting system containing the
genome and the expressome, defined as all the molecules (the 
transcriptome,
proteome, metabolome) produced according to the genetic program and
for which expression levels are regulated by genes and their epigenetic states 
(see Figure 1). Likewise the expression states of the genes
are regulated by the components of the expressome. For the sake of
model simplicity, but without loss of generality, we will specifically
talk about the genes expressing as, and being regulated by, proteins. 
However, other levels of description such as the metabolome could also be
viewed as relevant aspects of the expressome, similarly impacted by both 
endogenous and exogenous (environmental) factors.
We start from the organism in an initially ``normal'' state, i.e.
the state where all genes have youthful/healthy expression profiles. With
the passage of time, $t$, most of the genes,\textit{ $g(t)$} genes
in total, retain ``normal'' expression profiles, while a few genes,
$e_{g}(t)$ genes in total, subsequently become either physically
damaged or (epigenetically) dysregulated and represent a few ``defects''
or ``errors'' in the genetic program. The genes are translated into
the proteome, including its ``defects'', at a certain translation
rate \textit{p}. 

Next, we assume that the initial state of the genome is mostly stable, meaning that the
number of improperly expressed genes is small relative to the total
genome size, \textit{$e_{g}\ll G$} and, therefore, the number of
improperly produced proteins, $e_{p}\left(t\right)$, is also small.
This means that the interaction between the defects in the genome
and the proteome can be neglected. Therefore, the most general model
describing the dynamics of the interacting defects in the genome and
in the proteome can be described by the following equations:
\begin{equation}
\dot{e}_{p}=pe_{g}-ce_{p}+f_{p},\label{eq:basic_b}
\end{equation}

\begin{equation}
\dot{e}_{g}=\beta GKe_{p}-\delta e_{g}+Gf_{g},\label{eq:basic_M}
\end{equation}
where $\beta$ is the coupling rate constant characterizing the regulation
of the gene expression by the proteins. The constant $K$ is the average
number of genes regulated by any single protein and here represents
a simple measure of the overall connectivity of the genetic network
(See the section below on Strategies for stabilization of genetic networks). 
The constant $c$ reflects the combined
efficiency of proteolysis and heat shock response systems, mediating degradation and refolding 
of misfolded proteins, respectively, whereas
$\delta$ characterizes the DNA repair rate. Furthermore, the model
includes the force terms, $f_{p}(t)$ and $f_{g}(t)$, which characterize
the proteome and genome damage rates, respectively. The ``forces''
can represent any of a number of things, including oxidative stress
(metabolic), temperature, gamma-radiation (environmental), that are
imperfectly compensated by protective mechanisms. 

Eqs. (\ref{eq:basic_b}) and (\ref{eq:basic_M}) can only hold in
their simple linearized form if defects do not interfere with the
repair machinery or any other rare essential subsystems of the gene
network. As we will see below if, for example, a defect alters a DNA
repair system-associated gene or protein, the repair rate drops, the
system becomes unstable and may quickly diverge from its normal state
\cite{ohgkl1974maintenance,orgel2001maintenance}. To avoid the complications
from introducing such nonlinearities, we adopt a simple hypothesis
as to how defects in the evolving gene network could be responsible
for the demise of a cell or organism. Specifically, we assume that
mortality at any time is dependent on the probability of a defect
to ``land'' on and damage or dysregulate an essential gene. Because
any gene in the model can be dysregulated for a short time (brief
relative to lifespan) and then \textquotedblleft{}repaired\textquotedblright{},
a gene is considered essential if the disruption of its expression,
even for a limited time, is lethal to the cell in which it was disrupted. The suggested picture is quite
general, however, and is easily extended to multicellular/metazoan animals, since the stability boundaries
are the same \textendash{} only the exponents will be reduced because
some threshold fraction of cells must die (in some most-vulnerable
tissue compartment) to produce organism lethality. In this case the
population dynamics of a set of gene networks representing $N(t)$
organisms can be represented by: 
\begin{equation}
\dot{N}(t)=-M\left(t\right)N(t),\label{eq:basic_N}
\end{equation}
where $M(t)=\omega e_{g}/G$ is the mortality rate, proportional to
the fraction of mis-regulated genes, $e_{g}/G$, and $\omega$ is
an empirical factor, roughly a measure of the (small) fraction of
genes in the whole genome that are essential. The presented model is, obviously,
an extreme simplification. Nevertheless it can be rigorously derived
as a limiting case for the dynamics of a more complex genetic regulatory network,
where the number of dysregulated genes is sufficiently large
\cite{podolsky2013}.


Since aging is a slow process, Eqs. (\ref{eq:basic_b})-(\ref{eq:basic_N})
can be further simplified by neglecting second derivatives with respect
to time, 
\begin{equation}
\dot{M}(t)=\Lambda M(t)+\omega F(t),\label{eq:eq4M}
\end{equation}
where $F=f/(c+\delta)$ is a combined measure of genotoxic stress,
$f=\beta Kf_{p}+cf_{g}+\dot{f}_{g}$, and $\Lambda=(\beta pGK-c\delta)/(c+\delta)$
is the exponent characterizing genetic-network stability, which is precisely 
the propagation rate of gene-expression-level perturbations. In the
long run, stress levels can be averaged out and presumed to be time-independent,
$f_{p,g}\left(t\right)={\rm const}$, which yields the following expression
for the age-dependent mortality rate:

\begin{equation}
M(t)=\frac{\omega F}{\Lambda}(\exp\Lambda t-1).\label{eq:ML0-1}
\end{equation}
The nature of the solution is very different depending on the sign
of the exponent $\Lambda$. Whenever the combined efficiency of all repair systems 
is lower than a measure of the defect proliferation rate, 
\begin{equation}
R_{0}=\frac{\beta pG}{c\delta}K>1,\label{eq:R0}
\end{equation}
the gene network becomes inherently unstable, $\Lambda>0$, and the
number of regulatory errors (defects) grows exponentially, $M(t)\sim\exp(\Lambda t),$ which
is identical to the well-known Gompertz law \cite{gompertz1825nature}.
The average lifespan does not depend on the initial population size,
\begin{equation}
t_{le}\approx\frac{1}{\Lambda}\ln\left(1+\frac{1}{\gamma}+\sqrt{\frac{\pi}{2\gamma}}\right),\label{eq:tbar_unstable}
\end{equation}
but depends on both the exponent $\Lambda$ and on the genotoxic stress
level through the parameter $\gamma=\omega F/\Lambda^{2}$, which
is typically very small. Therefore in the limit, when the life expectancy is
large, the average lifespan can be approximated by $t_{le}\approx\Lambda^{-1}\log(1/\gamma)$.

A considerably more intriguing situation occurs when the genome is stable,
$\Lambda<0$ ($R_{0}<1$), and the gene network may remain stable
under reasonable stress conditions for a very long time. The fractions
of dysregulated genes and of misexpressed proteins
will then stabilize at constant levels, as will the mortality rate itself, $M_{\infty}=\omega F/|\Lambda|$. 
Constant mortality means that the population of animals dies off
exponentially rather than age-dependently: $N(t)=N_{0}\exp(-M_{\infty}t)$, which  is much slower
than the prediction from the Gomperz law. We believe that
the age-independent mortality observed in NMR experiments over a very
long lifespan \cite{buffenstein2005naked}, together with exceptional
stress resistance of NMR tissues \cite{lewis2012stress}, may be
manifestations of this stable scenario. We predict that the gene networks
of negligibly senescent animals are exceptionally robust, and the
number of dysregulated genes will scarcely change with age.
This argument is supported by the observations of \cite{kim2011genome},
in which the number of genes differentially expressed with age
was compared between NMR, mice and humans.

\begin{figure*}
\includegraphics[width=0.95\textwidth]{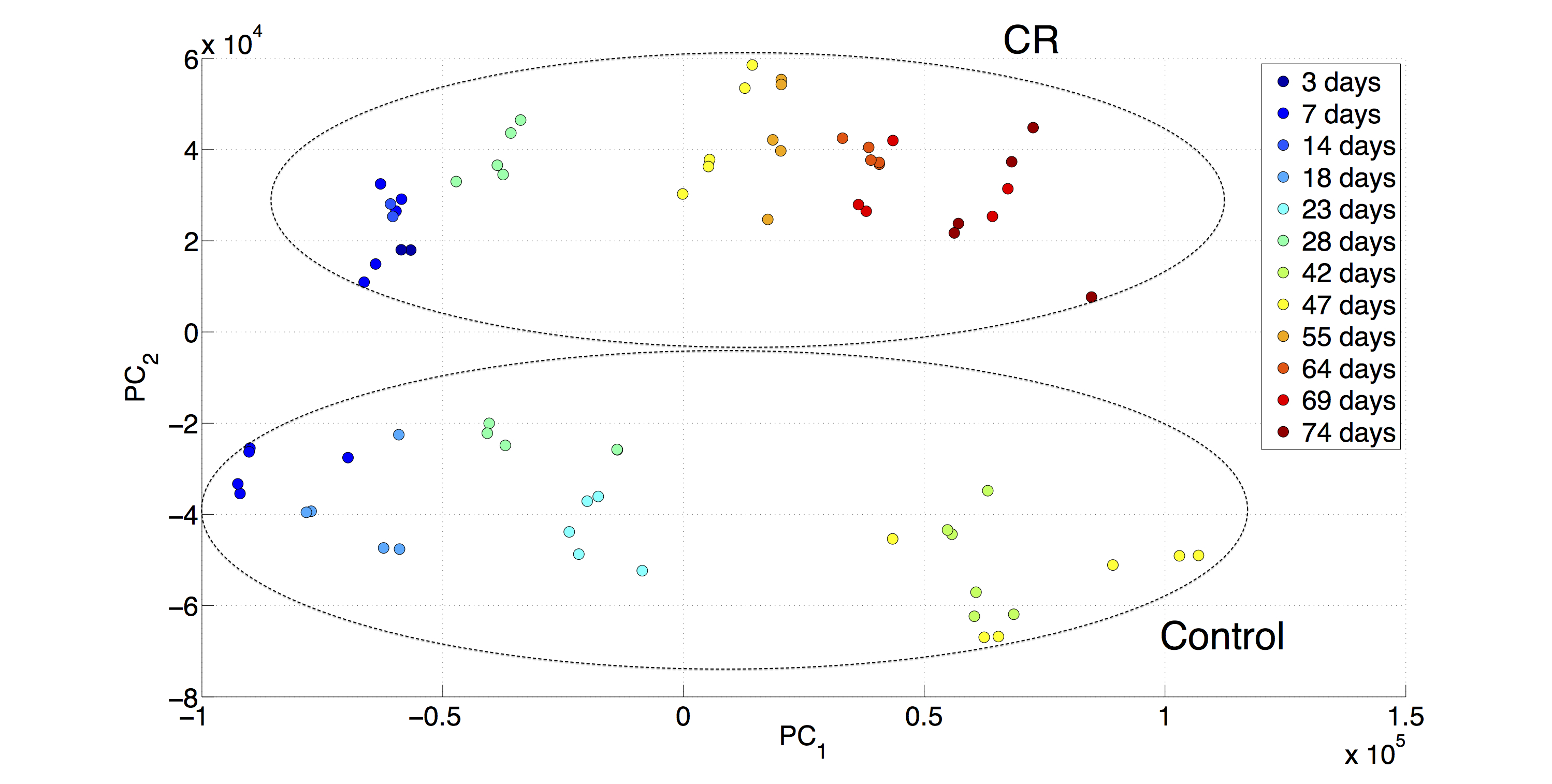}

\caption{Principal components analysis of gene expression profiles in aging
flies (data from \cite{pletcher2002genome}), fed on control (ad lib) and calorically
restricted diets. Every point represents a transcriptome for flies of a specific age and diet. As the animals
age, the genetic network accumulates regulation errors and the transcription
levels change in a single direction, up to a limit beyond which viability cannot be maintained.
\label{fig:Principal-component-analysis}}
\end{figure*}


\section*{Aging in a transcriptome of fruit flies}

The model summarized by Eqs. (\ref{eq:basic_N}) and (\ref{eq:eq4M}) predicts that the number of gene regulatory
errors will most often grow exponentially as animals age, as will the risk of death
(mortality). This means that gene expression levels (or metabolite levels, etc.) should change with age
and deviate from their healthy/youthful states. To test this prediction, we reanalyzed 
gene expression of fruitflies from the data of reference \cite{pletcher2002genome}. The 
measurements were performed at 6 adult ages, for two groups of
\textit{Drosophila melanogaster}: normally (``ad lib'') fed control flies and
calorically restricted (CR) flies. Figure \ref{fig:Principal-component-analysis} is a Principal Components (PC) analysis plot, in which
each point represents the state of gene expression for one combination of age and diet. The first/horizontal principal component, PC1, 
is the variance of an empirically derived cluster of 19 genes, which
strongly correlates with the age of adult animals in either of the two diet groups. 
Points on the extreme left correspond
to the youngest flies, and points for older age-groups are displaced progressively to the right. 

Thus, deviation of the gene expression profile from
the healthy young state increases with age, indicating that the number and extent of dysregulated genes increases 
along with mortality. However, no flies in either diet group were able to survive to the right of a certain boundary, beyond which the accumulation of gene-expression abnormalities becomes incompatible with survival of the organism.

Interestingly, the variance along the second principal component, PC2, creates a distinct separation between the ad lib (control) and CR-fed flies, based on expression variance in a different subset of 18 genes. This, of course, cannot be explained with the help of our basic model, but the proper generalization of that model is described in \cite{podolsky2013}.  
It is of great interest, nevertheless, that the entire range of variation in the PC1 dimension was less for CR flies than for control flies, despite spanning a much greater range of ages.

\section*{Genetic-network stability and stress resistance}

Extreme longevity has long been associated with exceptional resistance to a variety of stresses. And conversely, the decrease of stress resistance with age is one of the best-established indices of aging.  These phenomena can be explained using our simplified model. To
apply the model to experimental data, we reanalyzed data from reference \cite{parashar2008effects}, in which flies of varying age were exposed to radiation
during an exposure time $T_{1}\sim0.1 day$ and the median lethal dose
$LD_{50}$ was measured, corresponding to $50\%$ lethality of flies over a subsequent period of $T=2 days$. 
To model the experiment, we assumed that the animals
were subjected to an external genotoxic stress
at age $t=t_{0}$ for a duration $T$ that is small compared to the
lifespan of the animals. We solve Eqs. (\ref{eq:basic_N}), 
(\ref{eq:eq4M}) and  show that for ``normally''
aging animals, in which $\Lambda>0$, the dose required to produce $50\%$ mortality
decreases exponentially with age:
\begin{equation}
D_{50}\sim\frac{\ln2}{\omega T}-\frac{F_{0}}{\Lambda}\exp(\Lambda t_{0}),\label{eq: D_50}
\end{equation}
in qualitative agreement with the experimental data \cite{parashar2008effects}. We interpret this to mean
that stress resistance of ``normally'' aging animals decreases
progressively with age, and the average lifespan roughly corresponds
to the age at which endogenous (or common exogenous) stresses alone are sufficient to cause death. 
At late ages, $\Lambda t_{0}\gtrsim1$, the total accumulated
error load, $e_{g}(t)$, becomes so large that the linear model
defined by Eq. (\ref{eq:basic_N}) and (\ref{eq:eq4M}) fails. A more
comprehensive analysis shows that $LD_{50}$ remains positive, but shrinks exponentially
at the most advanced ages \cite{podolsky2013}. Eq.(\ref{eq: D_50})
also implies that measures of stress resistance should constitute robust biomarkers
of aging. 

We suggest that negligibly senescent organisms correspond to $\Lambda<0$.
That situation can be achieved only if the repair systems efficacies are
relatively strong. Moreover, the model predicts no further decrease with age 
in the lethal-stress dosage $LD_{50}$ beyond a threshold age, $|\Lambda|t_{0}\gtrsim1$.
Both predictions are well supported by experimental data. It is known
that negligibly senescent organisms are more stress resistant than
shorter-lived ones \cite{lewis2012stress}. For example, a comparison
of survival between negligibly senescent vs. short-lived clam tissues
treated with tert-butyl hydroperoxide showed much higher resistance to oxidative stress in long-lived clams
 \cite{ungvari2011extreme}.
Also Oxygen Radical Absorbance Capacity (ORAC) was measured in young
and old clams of both types. The experiment showed an age-related decline in ORAC for 
shorter-lived clams, whereas ORAC did not change with age in tissues of negligibly-senescent clams, which
is exactly what the model predicts for a negligibly-senescent animal. 

The proposed model may be considered as a general theory that subsumes
previous ``error catastrophe'' theories \cite{orgel2001maintenance,ohgkl1974maintenance} as special cases. 
It was long considered that error catastrophes had been disproved by the following experiment \cite{reis1976enzyme}. 
Drosophila adults in experimental groups were treated for $3-5$ days with a number of agents shown to 
increase misincorporation into protein or RNA, at doses leading to <20\% mortality. 
Although these treatments produced error rates much higher than were seen during aging of control 
flies, the misincorporation rates subsequently returned to control (pre-treatment) levels, 
and the average lifespans of survivors
were indistinguishable from controls \cite{reis1976enzyme}. Remarkably, this is exactly
the prediction from our model. Since mortality follows first-order kinetics
over time, the mortality increase from a stress depends only on the
current value of the stress. Thus, the solutions of Eqs. (\ref{eq:basic_b})-(\ref{eq:basic_N})
predict the same average lifespan of all animals surviving any stressful
treatment as for controls (although the average lifespan of the entire
treated group would obviously be shortened by inclusion of animals
that died during treatment). Of course, stresses that
induce appropriate defenses may cause hormesis, but this is beyond the intended scope
of our simplified linear model.

\section*{Strategies for stabilization of genetic networks}

Eq. (\ref{eq:R0}) implies many possibilities to stabilize a regulatory network 
and thus extend lifespan, as summarized in Table \ref{tab:GN-stabilization-strategies}.
According to Eq. (\ref{eq:tbar_unstable}),  a reduction
in genotoxic stress levels by isolation of genes from the
environment can protect from direct stresses but can only produce a
weak (logarithmic) increase in lifespan. Much stronger
effects could be achieved by interventions aimed at increasing
gene-network stability and hence reducing $\Lambda$. This can
be accomplished, for example, if genes and their epigenetic states were isolated from regulatory
signals, which equates to a decrease in $\beta$. 
This could be the reason why silencing of most signalling pathways -- which initially involve kinase cascades but mostly terminate in 
transcription factors -- accompanies extraordinary longevity and stress resistance in 
\textit{C. elegans} \cite{tazearslan2009positive, ayyadevara2008remarkable}.
Such defensive
strategies appear to have been utilized during the course of evolution, e.g., 
in protecting mitochondrial genes by their transfer
to the nuclear genome, and by establishment of the nuclear envelope, considered
a major factor leading to the emergence of multicellular life.

\begin{table*}
\begin{tabular}{|c|>{\centering}p{0.9\columnwidth}|}
\hline 
The model parameter & Biological embodiment\tabularnewline
\hline 
\hline 
Coupling rate, $\beta$ & Protective nuclear wall; transfer of mitochondrial genes to the nuclear
genome; inhibition of MGE(RT)\tabularnewline
\hline 
``Effective'' genome size, $G$ & Epigenetic inactivation of genes; tissue differentiation and specialization; temporal restriction of gene expression\tabularnewline
\hline 
Expressome (proteome, metabolome) turnover rate, $c$ & Turnover/repair of proteins and metabolites: their dilution via cell division, asymmetric division; chaperones, proteasomes, and autophagosomes; 
metabolome turnover/maintenance;
apoptosis (in multi-celled organisms)\tabularnewline
\hline 
DNA repair rate, $\delta$ & DNA repair; defenses against viruses and mobile genetic elements\tabularnewline

\hline 
Genotoxic stress level, $f$ & Isolation from environment; suppression of ROS; dietary preferences and avoidance of noxious biomaterials; development of nocireceptors and learned responses \tabularnewline
\hline 
\end{tabular}

\caption{Possible lifespan extension strategies with relation to the gene network
stability model parameters, and possible examples of their evolutionary deployment.\label{tab:GN-stabilization-strategies}}
\end{table*}

The relation between aging and the accumulation of defects is
not limited to defects in the proteome, but can readily be extended to defects in 
the metabolome, including lipid metabolites, and the glycome (both of which are 
largely regulated by proteins, and thus under the indirect control of genes). It also encompasses Mobile Genetic Elements (MGE),
genomic insertions of DNA ranging in size from hundreds to many thousands of basepairs, interspersed throughout
every mammalian genome (see a recent review \cite{moskalev2012role}). 
Although derived from ancient DNA and RNA viruses, very few of these elements encode an active ``transposase'', and
most have accumulated too many mutations to be transposed
(i.e., to jump within or between chromosomes) even in the presence of an active transposase,
which may be transiently provided by an exogenous retrovirus.  However, some MGEs are still mobile
and thus highly mutagenic, either passively or actively. Being essentially defects in the transcriptome
and interacting with the genome, this type of mutation fits our model
very well. MGEs are abundant, covering approximately 30\textendash{}50\%
of the human genome, and have been implicated as responsible for over 100
human genetic disorders including some cancers \cite{xing2007mobile,belancio2010somatic}.
Transposition of MGEs can be either cell-toxic or mutagenic, since they can disrupt essential genes 
or activate adjacent ones, so they may account
for a significant proportion of spontaneous genome damage. Considering
that MGEs evidently evolved from viruses, it is not surprising that
equations similar to Eqs. (\ref{eq:basic_b})-(\ref{eq:basic_N})
and the stability criterion similar to (\ref{eq:R0}) have been identified to describe 
infectious disease dynamics and stability of the virus-host system \cite{perelson2002modelling,baccam2006kinetics}.
The hypothesis that MGEs play a significant role in mutagenesis and
genomic instability predicts that lifespan may be increased by strategies aimed at reducing either
MGE transposition or recombination rates, $p$ or $\beta$, by inhibiting
the enzymes such as transposases and reverse transcriptases
(RTs), which mediate their mobility.  This prediction has been validated for at least some
species \cite{driver1994apparent}.

Turnover of the proteome or metabolome, $c$, and repair efficiency,
$\delta$, are factors shown to modulate lifespan in several species. 
Indeed, DNA repair pathways are encoded by hundreds of genes
\cite{milanowska2011repairtoire}, that are variously involved in
detection of DNA damage, enzymatic manipulation of damaged DNA, and
homologous recombination between DNA strands, often permitting complete restoration of
the original sequence even when portions have been lost from one chromatid or chromosome homolog \cite{moskalev2012role}. 
A somewhat more surprising
result is that both the proteome and genome repair rates, $c$ and
$\delta$ respectively, formally contribute to the result on an equal footing. This
means that increasing protein turnover may help protect against DNA
damage and\textit{ vice versa}. Increased protein turnover rates,
as can result from increased ubiquitin-proteasome activity, have been
demonstrated to result in increased longevity in yeast. Enhanced proteasomal
activity confers a 70\% increase in median and maximum replicative
lifespan, comparable to the longest lived single gene deletions identified,
and greater than the extension observed by deletion of TOR1 or 
over-expression of SIR2 \cite{kruegel2011elevated}.  

Recent data reveal that the NMR
has highly efficient protein degrading machinery and thereby maintains
high levels of protein quality control, constantly degrading misfolded
and damaged proteins, thus maintaining uniform steady-state levels throughout
life \cite{perez2009protein}. Our model predicts that
DNA repair efficiency, $\delta$, is as important for lifespan extension
as increases in the protein turnover rate, $c$. These NMR results are
interesting because their exceptional longevity occurs despite the presence
of chronic oxidative stress even at young ages \cite{rodriguez2012altered,lewis2012stress}.
Remarkably the stability of the proteome and its relationship to lifespan
were previously analyzed in a related context \cite{ryazanov2002protein}
with very similar conclusions. 

   Cell division is a trivial way
to rebuild the cell components and dilute expressome defects in half (symmetric
division) or more (asymmetric division), especially relevant to yeast and continuously dividing tissues. 
This simple dilution principle links the protein
turnover rate to cell division frequency, so that 
decreasing the protein turnover rate $c$, by inhibiting cell-cycling
or by other means, may be used as a gene-network destabilization strategy
for anti-cancer and antibiotic treatments \cite{dickson2009development}.

In differentiated organisms there appear even more ways to maintain
stability of the gene network. Metazoan (multi-tissued) animals have the ability
to eliminate cells that have sufficiently damaged or unstable genomes,
in a process called apoptosis, and replace them with healthy, stem cell-derived
cells. Thus the repair rates $c$ and $\delta$ also cover
the contributions of these apoptotic and regeneration pathways. Moreover,
as cells divide and differentiate to form new tissues and cell types,
resulting in many different epigenetically stable states of the same
genome, all of the model constants can also vary depending on the tissue
involved. Thus our model clearly permits the instability and aging rates of different
tissues to vary. This does not, however, alter the outcomes, which will reflect 
the vulnerability of the most unstable tissue on which animal survival depends -- 
most probably, the stem-cell subsystem corresponding to the most renewal-dependent tissue within the body.

According to the stability requirement (\ref{eq:R0}), large genomes
are difficult to maintain. There are multiple ways to increase 
system stability by keeping the size of the expressed genome under
control. One such strategy, clearly of ancient origin, is differentiation, 
wherein only a small fraction of the full genome is expressed by any one
cell type at any point in time. Another possible
way to regulate the stability of the genetic network is by modulating 
the degree of network connectivity. The robustness of a simulated
gene network with respect to external noise was recently shown to be associated
with the connectivity of the network \cite{peixoto2011emergence}.
This compares very well with our stability condition Eq.(\ref{eq:R0}),
which predicts that a gene network becomes stable if the characteristic
measure of network connectivity becomes sufficiently small: $K<K_{0}=c\delta/(p\beta G)$.
A recent study of long-lived \textit{Myotis brandtii} genomes found
a wide range of genetic abnormalities in the GH/IGF1 axis \cite{seim2013genome}.
Ablation of growth hormone signaling may produce a reduction of the effective
network connectivity $K$, which thus could be one of several possible explanations for 
the extreme life expectancy of long-lived bats.
All the above strategies are mathematically equivalent and exist in nature; indeed, there are cases in which 
multiple strategies are employed to increase lifespan \cite{seim2013genome}.

\section*{Conclusions and Prospects}

In summary, we have provided a mathematical explanation for the
dramatic variance in lifespans seen in the animal kingdom, relating this
variance to genetic-network stability and resistance to stresses.
We developed a model, which lets us  define Gompertzian or ``normal''
aging as an exponential accumulation of gene-regulation abnormalities
rooted in the inherent instability of gene networks occurring under
most common circumstances, and causing a progressive loss of stress resistance with age. This,
in turn, produces susceptibility to age-related diseases (or may even cause their onset), and leads to death of
an organism. This seems consistent with previously reported experimental
mouse data indicating that epigenetic dysregulation contributes far more 
(by up to two orders of magnitude) to the loss of gene-expression
integrity than somatic mutations alone \cite{li2012measuring}. 
On the other hand, gene-networks of animals with
better transcription fidelity or other mechanisms of genome-maintenance are not only more stress-resistant, but under specific conditions may become stable and produce negligible senescence phenotype
\cite{azpurua2013naked}.
An even more important corollary is that the gene networks of extremely
stress-resistant animals (extremophiles) can still be unstable, because network stability 
is determined by the value of the combined parameter $R_{0}$
given by Eq. (\ref{eq:R0}), and not directly by high $c$ and $\delta$ rates. 

The most important results of this study are Eq. (\ref{eq:eq4M}), phenomenologically describing aging of a gene
network, and the concept of fundamental genomic instability described by Eq. (\ref{eq:R0}).
 We show that the lifespan of a species is determined by the stability of its most vulnerable
gene network, coupled with its expressome in a realistic environment.
Mathematically, there exist two types of solutions to Eq. (\ref{eq:eq4M}), implying the possibility of both
stable and unstable phases. We posit that the stable phase corresponds
to negligible senescence and is robust with respect to environmental
or endogenous noise. Although unstable gene networks exponentially accumulate
gene expression errors and eventually disintegrate,
the growth exponent is nevertheless small. We believe that 
Eq. (\ref{eq:R0}) represents the same phase boundary previously identified 
in \cite{balleza2008critical}.
The critical dynamics of gene networks is a natural way to bridge
the ``hierarchy of scales'' problem, which is to explain how lifespans
can greatly exceed the time scales characterizing any of the constituent processes in living cells.
The genetic networks of most animals are inherently unstable and this leads to their 
aging. Since stabilization of gene networks can be favored in multiple
ways, further research has the clear potential to create novel
therapies to protect against the most morbid age-associated diseases, and perhaps even against aging itself.

\section*{Acknowledgments}

The authors are grateful to A. Tarkhov, M. Kholin and S. Filonov from
Quantum Pharmaceuticals, and also to Dr. D. Podolsky and M. Konovalenko, for comments
and stimulating discussions; and to Prof. A. Moskalev for extensive advice.

\bibliographystyle{naturemag}
\bibliography{Qrefs}

\end{document}